\documentclass[aps,pre,groupedaddress,showpacs,nofootinbib,showkeys,
twocolumn,amsmath,amsfonts]{revtex4-1}
\usepackage{graphicx}
\usepackage[utf8x]{inputenc} 
\usepackage{float}
\usepackage{bm}

\newcommand*\Laplace{\mathop{}\!\mathbin\bigtriangleup}
\newcommand{\vb}{\bar{v}}

\begin{document}


\title{Geometrical aspects in the analysis of microcanonical phase-transitions}


\author{Ghofrane Bel-Hadj-Aissa}
\affiliation{DSFTA, University of Siena, Via Roma 56, 53100 Siena, Italy}

\author{Matteo Gori}
\affiliation{Quantum Biology Lab, Howard University, 2400 6th St NW, Washington D.C. 20059, USA}

\author{Vittorio Penna}
%
\affiliation{Dipartimento di Fisica, Politecnico di Torino, Corso Duca degli Abruzzi 24,
I-10129 Torino, Italy}

\author{Giulio Pettini}
%
\affiliation{Dipartimento di Fisica Università di Firenze, and I.N.F.N., 
Sezione di Firenze, via G. Sansone 1, I-50019 Sesto Fiorentino, Italy}

\author{Roberto Franzosi}
\email[]{roberto.franzosi@ino.it}
\affiliation{QSTAR \& CNR - Istituto Nazionale di Ottica, Largo Enrico Fermi 2, I-50125 Firenze, Italy}


\date{\today}

\begin{abstract}
In the present work, we discuss how the functional form of thermodynamic observables
can be deduced from the geometric properties of subsets of phase space. 
 The geometric quantities taken into account are mainly extrinsic
curvatures of the energy level sets of the Hamiltonian of a system 
under investigation. In particular, it turns out that peculiar
behaviours of thermodynamic observables at a phase transition
point are rooted in more fundamental changes of the geometry
of the energy level sets in phase space. More specifically, we discuss
how microcanonical and geometrical descriptions of phase-transitions
are shaped in the special case of $\phi^4$ models with either nearest-neighbours and mean-field interactions.

\end{abstract}


\maketitle


\section{Introduction}

The microcanonical ensemble provides the statistical description of an isolated
system at equilibrium.
Within the microcanonical ensemble,
thermodynamic quantities, like temperature and specific heat, are derived from the entropy through
suitable thermodynamic relations.
The Boltzmann entropy is defined \cite{Huang_1987} as
\begin{equation}
S(E, \bm{\alpha}) = k_B \ln (\omega(E,\bm{\alpha}) \Delta)   \, ,
 \end{equation}
where $k_B$ is the Boltzmann constant, $\omega(E,\bm{\alpha})$ is the density of
microstates per unit energy interval,  $E$ is the total energy, $\bm{\alpha} =(\alpha_1,\ldots\alpha_m)$ summarizes external parameters as,
for instance, the volume, and $\Delta$ is a constant with the dimension of an energy. 
In the following we assume units such that $k_B =1$.
We have to recall that, in point of fact, recently \cite{Dunkel2013} it has been argued that only the Gibbs entropy yields
a consistent thermodynamics, whereas the microcanonical statistical mechanics founded on the Boltzmann entropy,
would unveil some issues. We do not share the point of view of these authors as we have clarified
in Refs. \cite{Buonsante_AoP_2016, Buonsante_2017,Franzosi_PA_2018} where we have shown that the Boltzmann
entropy provides a consistent description of the microcanonical ensemble.

We have already mentioned that the entropy $S$ constitutes the fundamental
thermodynamic potential in the microcanonical ensemble. Indeed, from the entropy $S$
of a given system, secondary thermodynamic observables, such as the absolute temperature $T$ or the pressure $p$,
are derived by differentiation of $S$ with respect to the parameters $\{E, \bm{\alpha}\}$.
Let us denote the partial derivatives with respect to $E$ by a prime, in this case the
temperature $T$ is
\begin{equation}
{T} = \left( \dfrac{\partial S}{\partial E} \right)^{-1}  = \dfrac{\omega}{\omega^\prime}\, .
\label{eq:T0}
\end{equation}
Similarly, for the specific heat we have 
\begin{equation}
C_v = - \dfrac{\left(\dfrac{\partial S}{\partial E} \right)^2}
{\left( \dfrac{\partial^2 S}{\partial E^2}\right)} = 
\dfrac{(\omega^\prime/\omega)^2}{(\omega^\prime/\omega)^2 - 
\omega^{\prime \prime}/\omega}\ .
\label{eq:cv0}
\end{equation}

In the following, we will derive the explicit formulas for several thermodynamic quantities in the
case of a generic Hamiltonian system. Indeed, in this case, the geometric structure related
to the dynamics allows one to derive explicit formulas that can be used in numerical simulations of a 
Hamiltonian flow to work out thermodynamic observables through time averages.
Furthermore, we consider two lattice models for which we specify such formulas and for which we perform
numerical simulations in order to exemplify the method. In addition, we consider several geometric
quantities that seem to clearly detect the phase transition point.

\section{Geometric microcanonical thermodynamics}
The geometric microcanonical formalism in the general case of an
Hamiltonian system is derived as follows.
Consider a generic classical many-particle system described by an autonomous
Hamiltonian $H (x_1,\ldots, x_{N} )$ depending on $N$ canonical coordinates, $\bm{x}=(\bm{p},\bm{q})$
in which the energy is the only first integral of motion.
In this case the Boltzmann entropy reads
\begin{equation}
S(E) =  \ln\! \int \! d^{N}\bm{x} \,
\delta(E - {H}(\bm{x})) \, ,  
\label{entropyb}
\end{equation}
where $\delta$ is the Dirac function.

In accordance with the conservation of energy, during its evolution in time the representative point of the 
system moves on a given energy-level set.
The Liouville theorem shows that the measure of the Euclidean volume is
preserved by the dynamics. Consistently, the invariant measure
$\mu$ induced on each energy level set $\Sigma_E$, of energy
$E$, is given by
\cite{khinchin_1949}
\begin{equation}
{ d \mu} = \dfrac{d \Sigma}{\Vert \nabla H \Vert} \, ,
\label{hs-measure}
\end{equation}
where $d \Sigma$ is the Euclidean measure induced on $\Sigma_E$, and
$\Vert \cdot \Vert $ is the Euclidean norm. 
Remarkably, the density of microstates corresponding to $E$ just depends
on such invariant measure as
\begin{equation}
\omega (E) = \int_{\Sigma_E} d \mu \, .
\label{omega}
\end{equation}
Since temperature, specific heat and other thermodynamic quantities depend
on the derivatives of the microcanonical entropy with respect to the energy,
they can be measured by averages of the form
\begin{equation}
\langle \Phi \rangle = \dfrac{\int_{\Sigma_E} d \mu \, \Phi }
{\int_{\Sigma_E} d \mu } \, ,
\label{phimu}
\end{equation}
where $\Phi$ stands for any function of interest.
The latter fact has been proven by Rugh \cite{RughPRL97} in the case of
Hamiltonian systems for which the energy is the only conserved quantity
and, in Ref. \cite{Franzosi_JSP11} and Ref.
\cite{Franzosi_PRE12} for the case of two and $k\in \mathbb{N}$, $k>2$
conserved quantities, respectively. The formalism derived in
\cite{Franzosi_JSP11} has been successfully adopted for the study of the
microcanonical thermodynamics of systems describing Bose-Einstein condensates
in optical lattices
\cite{Franzosi_Nonlinearity_24,Iubini_2013,Buonsante_AoP_2016,Buonsante_2017}
for which there exist two conserved quantities.

The two $\phi^4$ models considered in the following have just one conserved
quantity, that is the total energy, therefore from now on we will limit
our discussion to the the simpler case studied in Ref. \cite{RughPRL97}.
The geometric key tool in this case is the Federer-Laurence derivation
formula 
\cite{Federer_1969,Laurence_1989}
\begin{equation}
\partial^k (\int_{\Sigma_E}  \psi
 d \Sigma )/\partial E^k  = \int_{\Sigma_E} A^k\left( \psi\right)
 d \Sigma \, ,
\label{eq:F-L-formula}
\end{equation}
where 
\begin{equation}
 A(\bullet) = \frac{1}{\Vert\nabla H\Vert}\nabla
 \left(\frac{ \nabla H }{\Vert\nabla H \Vert } \bullet\right) \, .
\label{eq:A} 
\end{equation}

By using this formula in the inverse temperature definition
\begin{equation}
\dfrac{1}{T} = \dfrac{\partial S}{\partial E}
\label{Temperat}
\end{equation}
one obtains

\begin{equation}
\beta =\dfrac{\omega^\prime}{\omega} = \langle \Phi_1 \rangle \, ,
\label{betarugh}
\end{equation}
where $\beta = 1/T$,
$\omega^\prime = \frac{\partial \omega}{\partial E}$ and
\begin{equation}
\Phi_1 = 
\nabla \cdot \left({\nabla H}/{\Vert \nabla H \Vert^2} \Phi_0 \right)  \, ,
\label{Phi1}
\end{equation}
with $\Phi_0 = 1$.

It is worth mentioning here that in a recent paper \cite{Franzosi_PA_2018},
one of the present authors has suggested using the surface entropy, that is
the logarithm of the area of the constant energy hypersurfaces in the phase
space, as the definition for the thermodynamic microcanonical entropy,
in place of the standard definition \eqref{entropyb}. 
Besides the fact that the surface entropy has properties which make it an
attractive definition for small systems \cite{Franzosi_JSTAT_2019}, from a
geometric point of view, the inverse temperature $\beta_s$ derived from
the surface entropy is linked to the mean curvature of the hypersurface
$H(\bm{x}) = E$, that is with a geometric quantity.
In fact, in \cite{Franzosi_JSTAT_2019} it is shown that in the case of the
surface entropy the inverse temperature results
\begin{equation}
\beta_s = 
\dfrac{  \int_{\Sigma_E} M(x) m^{N-1}(\Sigma_E) }
{ \int_{\Sigma_E} m^{N-1}(\Sigma_E) } \, , 
\label{meancurvature}
\end{equation}
where
\begin{equation}
M(x) = \dfrac{1}{\Vert \nabla H \Vert}
\nabla \cdot \left(
\dfrac{\nabla H}{\Vert \nabla H \Vert}
\right)
\label{meancurvature1}
\end{equation}
is the local mean curvature divided by $\Vert \nabla H \Vert$.

Coming back to the standard entropy definition \eqref{entropyb},
after the Federer-Laurence derivation formula
\eqref{eq:F-L-formula} we get
\begin{equation}
\dfrac{\omega^{\prime \prime}}{\omega} = \langle \Phi_2 \rangle
\, ,
\label{omega2}
\end{equation}
where
\begin{equation}
\Phi_2 =  
\nabla \cdot
 (\nabla H/\Vert \nabla H \Vert^2 \Phi_1) \, ,
\end{equation}
thus the specific heat \eqref{eq:cv0} results
\begin{equation}
C_v = 
\dfrac{ \langle \Phi_1 \rangle^2}{ \langle \Phi_1 \rangle^2 - 
 \langle \Phi_2 \rangle}\, .
\label{eq:cv}
\end{equation}

More generally, the derivative of order $k$ is obtained by a recursion
according to the relation
\begin{equation}
\dfrac{1}{\omega} \frac{\partial^{k} \omega}{ \partial E^{k}}=
\langle \Phi_{k} \rangle
\label{eq:genOk}
\end{equation}
where
\begin{equation}
\Phi_{k} = \nabla \cdot
 (\nabla H/\Vert \nabla H \Vert^2 \Phi_{k-1}) \, .
\label{eq:recPhi}
\end{equation}
We will discuss in the following the relevance of the behaviour of
the second derivative of the entropy with respect to the energy density $E/N$.
The latter quantity is expressed in terms of the averages of $\Phi_1$ and
$\Phi_2$ according to the following equation
\begin{equation}
\left( \dfrac{\partial^2 S}{\partial E^2}\right) = \langle \Phi_2 \rangle  - 
\langle \Phi_1 \rangle^2
\label{d2sf}
\end{equation}

In the following we will report the microcanonical inverse temperature $\beta(E)$ and the
specific heat $C_v(E)$ by  time averages of the relevant functions
$\Phi_1$ and $\Phi_2$. In fact, under the hypothesis of ergodicity, the microcanonical 
averages of each observable $\Phi$ can be equivalently measured along the
dynamics according to
\begin{equation}
\langle \Phi \rangle = \lim_{\tau \to\infty} \dfrac{1}{\tau} \int_0^\tau dt \, \Phi(t) \, ,
\label{timeav}
\end{equation}
equivalent to an average on $\Sigma_E$ as in Eq. \eqref{phimu}.

The explicit form for the function $\Phi_1$ is
\begin{equation}
\Phi_1 = \dfrac{\Laplace H}{\Vert \nabla H \Vert^2} - 2 
\dfrac{\nabla H \cdot {\cal H} \cdot \nabla H }{\Vert \nabla H \Vert^4} \, ,
\label{Phi1}
\end{equation}
where ${\cal H}$ is the Hessian matrix of the Hamiltonian function, 
whereas $\Phi_2$ is a little bit more complicated
\begin{equation}
\Phi_2 = \Phi_1^2 + \dfrac{\nabla H \cdot \nabla (\Phi_1)}{\Vert \nabla H \Vert^2} \, .
\label{Phi2}
\end{equation}

\subsection{Phase transitions in the microcanonical ensemble}
The inequivalence of statistical ensembles in presence of long-range interactions, and Molecular Dynamics studies of energy conserving systems have motivated several investigations of the microcanonical description of phase transitions \cite{gross,micro1,micro2,micro3,micro4,PRL-Bachmann}. In particular, we emphasize a recent and very interesting proposal in Ref. \cite{PRL-Bachmann} which proves very effective to interpret the outcomes of numerical simulations, as it will be seen in the following.  A complementary viewpoint  \textit{\`a la Ehrenfest} has been heuristically put forward in \cite{Pettini2019}. This proceeds from the fact that the natural counterpart of microscopic Hamiltonian dynamics is the microcanonical ensemble where, as we have already recalled above, the relevant thermodynamic potential is entropy. From the latter, one derives the specific heat \eqref{eq:cv0},  
where we see that $C_v$ can diverge only as a consequence of the vanishing of $({\partial^2 S}/{\partial E^2})$ which a-priori has nothing to do with a loss of analyticity of $S(E)$. This disagrees with Ehrenfest's classification of phase transitions in the canonical ensemble, associated with a loss of analyticity of Helmholtz free energy, and thus also of the entropy. As is well known, the identification of a phase transition with an analyticity loss of a thermodynamic potential (in the gran-canonical ensemble) is rigorously stated by the Yang-Lee theorem. 

Coming to the microcanonical ensemble, for standard Hamiltonian systems
(i.e. quadratic in the momenta) the relevant information is carried by the
configurational microcanonical entropy 
\begin{equation}
      S_n(\vb) =\frac{1}{n} \log{ \int dq_1\cdots dq_n\ \delta
[V_n (q_1,\dots, q_n) - v] ~, \label{pallaM}} \ \nonumber
%
\end{equation}
where $\vb = v/n$ is the potential energy per degree of freedom, $\delta[\cdot]$ is the Dirac function, $S_n(\vb)$ is related to the configurational canonical free energy
      \[
    f_n(\beta)=   \frac{1}{n} \log \int dq_1\dots dq_n\ e^{-\beta V_n(q_1,\dots, q_n)}\nonumber
    \]
for any $n\in{\mathbb N}$, $\vb\in{\mathbb R}$, and $\beta \in{\mathbb R}$,
     through the Legendre transform 
    \begin{eqnarray}
     - f_n(\beta) =  \beta \cdot \vb_n -  S_n(\vb_n)\, ,
    \label{legendre-tras}
    \end{eqnarray}
where the inverse of the configurational temperature $T(v)$  is given by $\beta_N(\vb)= {\partial S_N(\vb)}/{\partial \vb}$.

Then consider the function $\phi(\vb)=f_n[\beta(\vb)]$, from $\phi^\prime(\vb) = -\vb\  [d \beta_n(\vb)/d\vb]$
we see that if $\beta_n(\vb)\in{\cal C}^k(\mathbb R)$ then also $\phi(\vb)\in{\cal C}^k(\mathbb R)$ which in turn
means $S_n(\vb)\in{\cal C}^{k+1}(\mathbb R)$ while $f_n(\beta)\in{\cal C}^k(\mathbb R)$. 
Hence, if the functions $\{S_n(\vb)\}_{n\in{\mathbb N}}$ are convex, thus ensuring the existence of the above Legendre transform, and if in the $n\to\infty$ limit it is $f_\infty(\beta)\in{\cal C}^0(\mathbb R)$ then $S_\infty(\vb)\in{\cal C}^1(\mathbb R)$, and if 
$f_\infty(\beta)\in{\cal C}^1(\mathbb R)$ then $S_\infty(\vb)\in{\cal C}^2(\mathbb R)$. {{So far we have seen that, generically (that is apart from any possible counterexample), if $f_n(\beta)\in{\cal C}^k(\mathbb R)$ then $S_n(\vb)\in{\cal C}^{k+1}(\mathbb R)$. This all what is needed to \textit{heuristically} proceed to a classification of phase 
transitions {\it \`a} {\it la} Ehrenfest in the present microcanonical configurational context. 
Indeed, in the original definition, Ehrenfest associates a first or second-order phase transition with a discontinuity in the first or second derivatives of $f_\infty(\beta)$, respectively.
By analogy, we associate a first (second) order phase transition with a discontinuity
of the second (third) derivative of the entropy $ S_\infty(\vb)$. 
It is worth emphasizing that this definition of the order of a phase transition  is given regardless of the existence of the  Legendre transform. Indeed, the latter is very often unachievable in the presence of first-order phase transitions which bring about a kink-shaped entropy as a function of the energy \cite{gross}. 
Therefore, rigorously, the definition that we are putting forward does not stem either mathematically and logically from the original Ehrenfest classification. 

This entropy-based classification of phase-transitions  {\it \`a} {\it la} Ehrenfest,
although to some extent arbitrary, has a \textit{heuristic} motivation.
Besides, it does not suffer any  longer the difficulty arising
in the framework of canonical ensemble where a divergent specific heat can exist also in presence of a second-order phase transition, 
as is the case of the Ising model in dimension two.
This classification is useful also in case of ensemble non-equivalence when only the microcanonical description is the correct one. 

The usefulness of this classification has to be confirmed against
practical examples beyond Ref. \cite{Pettini2019}.

\section{The models}
The $\phi^4$ models are defined by the Hamiltonian
\begin{equation}
H = \sum_{\bf j} \dfrac{1}{2} \pi^2_{\bf j}  + V(\phi)
\label{Hphi4}
\end{equation}
where
\begin{equation}
\!\!V(\phi) =\sum_{\bf j} \left[ 
 \dfrac{\lambda}{4!} \phi^4_{\bf j} - \dfrac{\mu^2}{2}\phi^2_{\bf j} +
\dfrac{J}{D} \sum_{{\bf k}\in I({\bf j})} (\phi_{\bf j} - \phi_{\bf k})^2
 \right] \, ,
\label{Vphi4}
\end{equation}
$\pi_{\bf j}$ is the conjugate momentum of the variable $\phi_{\bf j}$
that defines the position of the ${\bf j}^{th}$ particle.
In the case of the two dimensional model, 
${\bf j} = (j_1,j_2)$ denotes a site of a two dimensional lattice,
the number of nearest neighbours is $D=4$ and
$I({\bf j})$ are the nearest neighbour lattice sites of the
${\bf j}^{th}$ site. The coordinates of the sites are integer numbers
$j_k =1,\ldots,N_k$, $k=1,2$, so that the total number of sites
in the lattice is $N=N_1\,N_2$. Furthermore periodic boundary conditions
are assumed. In the case of the mean-field model
${\bf j}=1,\ldots,N$ denotes the indices of the $2N$ canonical coordinates
of the system, $D=N-1$ and $I({\bf j}) = 1,\ldots,N$.
The Hamiltonian equations of motion read
\begin{equation}
\begin{array}{l}
\dot{\phi}_{\bf j} = \pi_{\bf j} \, , \\
\dot{\pi}_{\bf j} =  - \dfrac{\partial V}{\partial \phi_{\bf j}}
\end{array} \, .
\label{eq:motion}
\end{equation}
The local potential displays a double-well shape whose minima are located 
at $\pm \sqrt{{3! \mu^2}/{\lambda}}$ and to which it corresponds 
the ground-state energy per particle $e_0 = - 3! \mu^4/(2 \lambda)$.
At low-energies the system is dominated by an ordered phase where the time 
averages of the local fields are not vanishing. By increasing the system energy 
the local $\mathbb{Z}_2$ symmetry is restored and the averages of the local-
fields are zero.

Naturally, the explicit form for the geometric quantities entering in
\begin{equation}
\Phi_1 = \dfrac{\Laplace H}{\Vert \nabla H \Vert^2} - 
2 \dfrac{\nabla H \cdot {\cal H} \cdot \nabla H}{\Vert \nabla H \Vert^4}
\, ,
\label{Phi1e}
\end{equation}
and
\begin{equation}
\Phi_2 = \Phi_1^2 + \dfrac{\nabla H}{\Vert \nabla H \Vert^2}
\cdot \nabla (\Phi_1) \, ,
\label{Phi2e}
\end{equation}
where 
\begin{equation}
\begin{split}
 \dfrac{\nabla H}{\Vert \nabla H \Vert^2} &
\cdot \nabla (\Phi_1) =\\  
\sum_{{\bf j} {\bf k}} \dfrac{\partial_{\bf j} H \partial^3_{{\bf j}{\bf k}{\bf k}}
H} 
{\Vert \nabla H \Vert^4} &
- 2 
\sum_{{\bf j} {\bf k} {\bf r}} \dfrac{\partial_{\bf j} H 
\partial_{\bf k} H \partial_{\bf r} H 
\partial^3_{{\bf j}{\bf k}{\bf r}} H}{\Vert \nabla H \Vert^6} + \\
-4 \dfrac{\nabla H \cdot {\cal H} \cdot {\cal H} \cdot \nabla H}
{\Vert \nabla H \Vert^6}& - 2 
\dfrac{\nabla H \cdot  {\cal H} \cdot \nabla H}
{\Vert \nabla H \Vert^4} \times \\ &
\left(\Phi_1 - 2 \dfrac{\nabla H \cdot  {\cal H} \cdot \nabla H}
{\Vert \nabla H \Vert^4} \right) \, ,
\label{Phi2S}
\end{split}
\end{equation}
depend on the details of the Hamiltonian
of each model.
Therefore, in the following we will consider these two cases
separately and we will set 
\[
\nabla \equiv 
\left(
\begin{array}{c} \vdots \\
\partial_{\pi_{\bf j}}\\
\vdots \\
 \partial_{\phi_{\bf j}} \\
\vdots 
\end{array}
\right) \, .
\]
 
\subsection{\emph{$2$-d $\phi^4$ model.} }
In the case of the two dimensional model we have
\begin{equation}
\Laplace H = N (1+4J-\mu^2) + \dfrac{\lambda}{2!} \Vert \phi \Vert^2 \, ,
\label{lapla} 
\end{equation}
where  $\Vert \phi \Vert = \sqrt{\sum_{\bf j} \phi^2_{\bf j}}$.
In addition it results 
\begin{equation}
\Vert \nabla H \Vert = \sqrt{2 K + \Vert \nabla V \Vert^2} \, ,
\label{nnablaH}
\end{equation}
where $K$ stands for the total kinetic energy
$K = \sum_{\bf j}  \pi^2_{\bf j}/2$ and
\begin{equation}
\nabla_{\bf k} V = \dfrac{\lambda}{3!} \phi^3_{\bf k} +
(4 J -\mu^2)  \phi_{\bf k} -J \sum_{{\bf j}\in I({\bf k})}
  \phi_{\bf j} \, .
\label{nablaV}
\end{equation}
The Hessian matrix of the Hamiltonian function is
\begin{equation}
{\cal H} =
\left(
\begin{array}{cc}
\mathbb{I} & 0 \\
0 & {\cal H}_V
\end{array}
\right) \, ,
\label{hess}
\end{equation}
where the entries of the Hessian matrix ${\cal H}_V$ of the
potential function $V$ result
\[
({\cal H}_V)_{{\bf i}{\bf j}} = \partial^2_{{\bf i}{\bf j}} V =
\left(\dfrac{\lambda}{2!} \phi^2_{\bf j} + 4 J - \mu^2 \right)
\delta_{{\bf i},{\bf j}} -J \delta_{{\bf j},I({\bf i})} \, .
\] 
Finally, it is
\[
\partial^3_{{\bf i}{\bf j}{\bf k}} V =  \lambda
\delta_{{\bf i},{\bf j}} \delta_{{\bf j},{\bf k}} \phi_{\bf j} \, .
\] 
 
\subsection{\emph{Mean-field $\phi^4$ model.} }
The analogous quantities for the case of the mean-field model are the
following.
$\Laplace H$ has the same form of \eqref{lapla}, whereas
\begin{equation}
\nabla_{\bf k} V = \dfrac{\lambda}{3!} \phi^3_{\bf k} +
\left[ 4 J\dfrac{N}{N-1} -\mu^2\right]  \phi_{\bf k} 
-\dfrac{4 J}{N-1} {\cal M} \, ,
\label{nablaV}
\end{equation}
where we have introduced the total magnetization 
\begin{equation}
{\cal M} =  \vert \sum_{{\bf j}}
  \phi_{\bf j} \vert \, .
\label{M}
\end{equation}
In this case the Hessian matrix ${\cal H}_V$ of the
potential function $V$ is
\[
({\cal H}_V)_{{\bf i}{\bf j}} = \partial^2_{{\bf i}{\bf j}} V =
\left[ \dfrac{\lambda}{2!} \phi^2_{\bf j} + 4 J 
\dfrac{N}{N-1} - \mu^2 \right]
\delta_{{\bf i},{\bf j}} -\dfrac{4J}{N-1} \, ,
\] 
and $\partial^3_{{\bf i}{\bf j}{\bf k}} V$ has the same
form of the $2-$d case. 

\section{ Numerical results}
We have investigated the microcanical thermodynamics of these two
systems by measuring some geometric quantities as illustrated
above which are relevant to catch the thermodynamical properties
of these models at their respective phase transition points.
Thus, we have numerically integrated the equations of motion
\eqref{eq:motion} of both models, by using a third
order symplectic algorithm \cite{lapo} and starting from random initial
conditions corresponding to different values of the total
energy $E$.
In such a way, we have measured - along the dynamics - the time averages
of the  quantities $\Phi_1$ and $\Phi_2$ for several values of
the total energy $E$, according to Eq. \eqref{timeav}.
From the time averages $\langle \Phi_1 \rangle (E)$ and
$\langle \Phi_2 \rangle(E)$ by means of Eqs. \eqref{betarugh} and
\eqref{eq:cv}, we have derived the caloric curve
$T(E)$ and the specific heat $C_v(E)$ of
the two models. In addition to the thermodynamic
quantities, we have measured geometric quantities as the 
average of the Ricci curvature
$K_R(q,\dot{q})$
(see App. \ref{Eisenhart_metric} for details). 
The main outcome of our
analysis is the better effectiveness of the geometric indicators as
phase-transitions detectors with respect to the traditional thermodynamic
indicators, with the exception of the order parameter.
In a recent paper
 \cite{Pettini2019},
by resorting to geometric indicators, it has been possible to unambiguously characterize and explain the phenomenology of
a system that undergoes a thermodynamic phase transition in the
absence of a global symmetry-breaking and thus in the absence of
an order parameter.

\subsection{ $2$-d $\phi^4$ model}
In this section, we analyse the results of the simulations performed
in the case of the $2d$ $\phi^4$ model. In two dimension the
$\phi^4$ model undergoes a second-order phase transition which
is detected by a bifurcation of the order parameter. This is clearly seen in Fig. \ref{order_e_2d} which reports the typical behaviour of the order parameter
$M=\langle {\cal M}  \rangle/N$,  the average of the total 
magnetization ${\cal M}$ defined in \eqref{M}, as a function of the energy
density $E/N$, behavior typical of a  second-order phase transition.
\begin{figure}[h]
 \includegraphics[height=5.5cm]{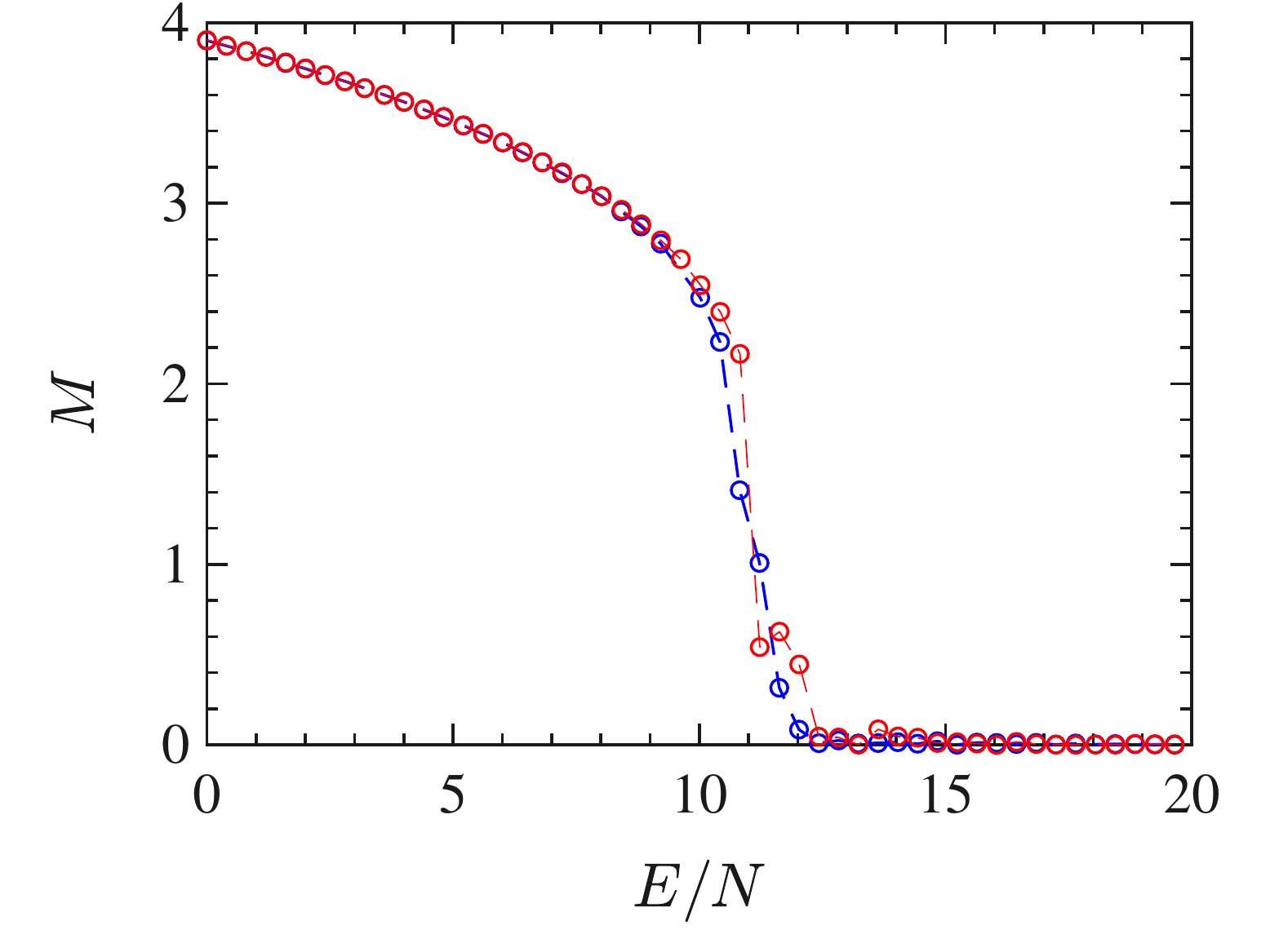}
\caption{\label{order_e_2d} The figure shows the plot
of the quantity order parameter $M$
vs the energy density $E/N$ for $128 \times 128$ particles
(blue circles) and $48\times 48$ particles
(red circles).}
\end{figure} 
Figure \ref{order_e_2d} allows one to determine the critical energy density
$\epsilon_c$ of the phase-transition, which is found to be 
$\epsilon_c \approx 11.1$.
\begin{figure}[h!]
 \includegraphics[height=5.5cm]{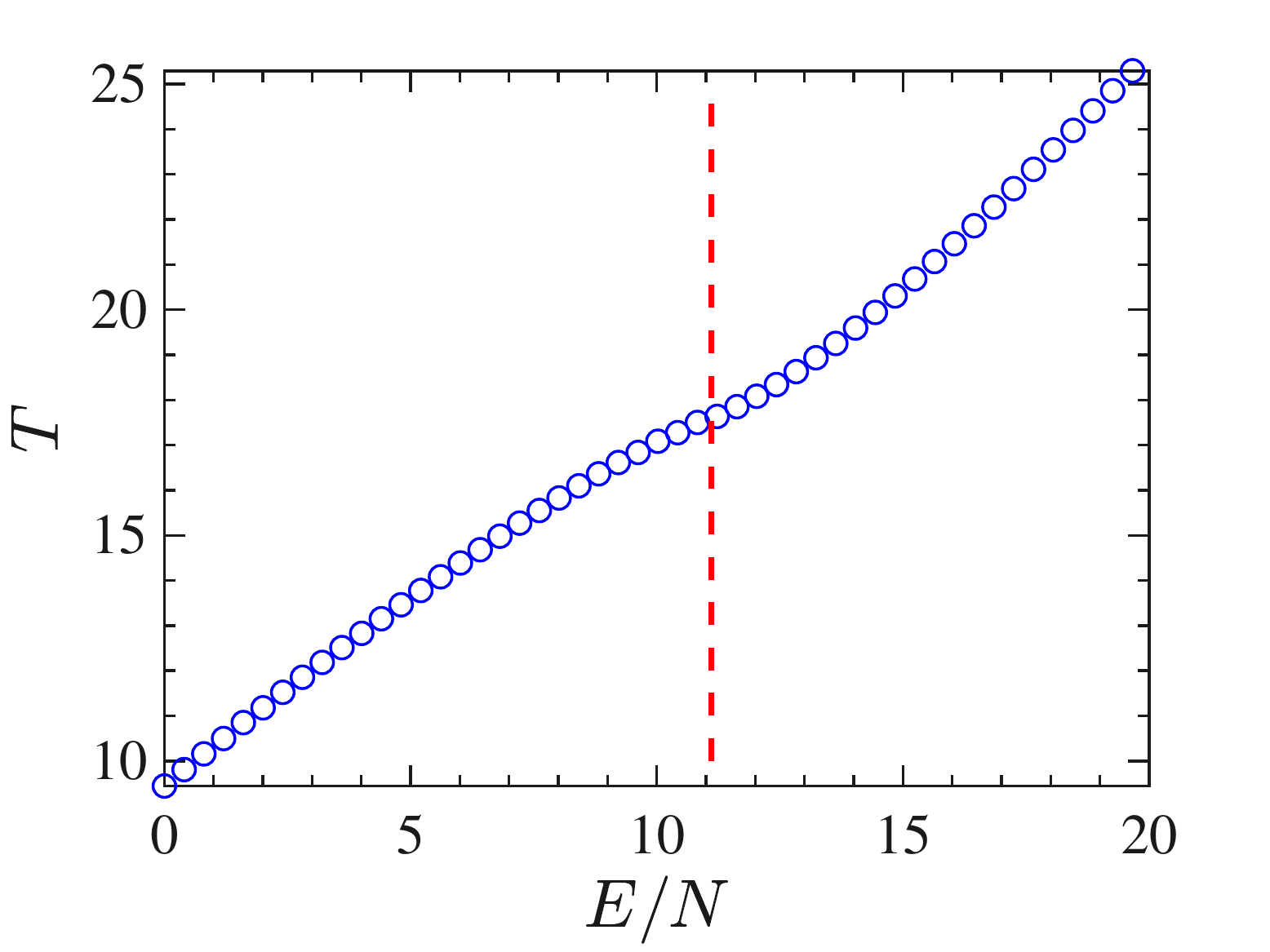}
\caption{\label{e_T_2d} The figure reports the temperature derived by means
of time averages 
of $\Phi_1$ whence $T=1/\Phi_1$ according to Eqs. \eqref{betarugh} and \eqref{timeav},
as a function of the energy density $E/N$ for the $2-d$ $\phi^4$ model
for  $128 \times 128$ particles.}
\end{figure} 
Another typical signature of a phase transition is provided by the shape of the caloric curve $T(E)$, i.e. 
the temperature as a function of the energy. In the case of the $2d$ $\phi^4$
model, we have derived such a curve by time-averaging $\Phi_1$, along
with the dynamics, for different initial conditions corresponding to several
energy densities. The caloric curve derived by $1/\langle \Phi_1 \rangle$,
according to Eq. \eqref{betarugh}, is reported in Fig.  \ref{e_T_2d}.
In the case of the $2d$ $\phi^4$ model, the caloric curve $T(E/N)$ displays an inflection point just at the critical energy density value
identified  by the bifurcation point of the order parameter -  highlighted with the vertical dfashed line in Fig.  \ref{e_T_2d} - and this is in
perfect agreement with the proposition put forward by Bachmann in Refs. \cite{Bachman_book,PRL-Bachmann}.
Through time averages of $\Phi_1$ computed along with the numeric
integration of the equations of motion for different initial conditions,  
we have derived the curve of the inverse temperature $\beta$ as a function
of $E/N$. Fig. \ref{beta_e_2d} shows this curve for the $2d$ $\phi^4$ model.
Also in the case of $\beta(E/N)$, the transition point $E_c/N =\epsilon_c$
(located by the
dashed vertical line in the same figure) corresponds to an inflection point of this curve.
\begin{figure}[h]
 \includegraphics[height=5.5cm]{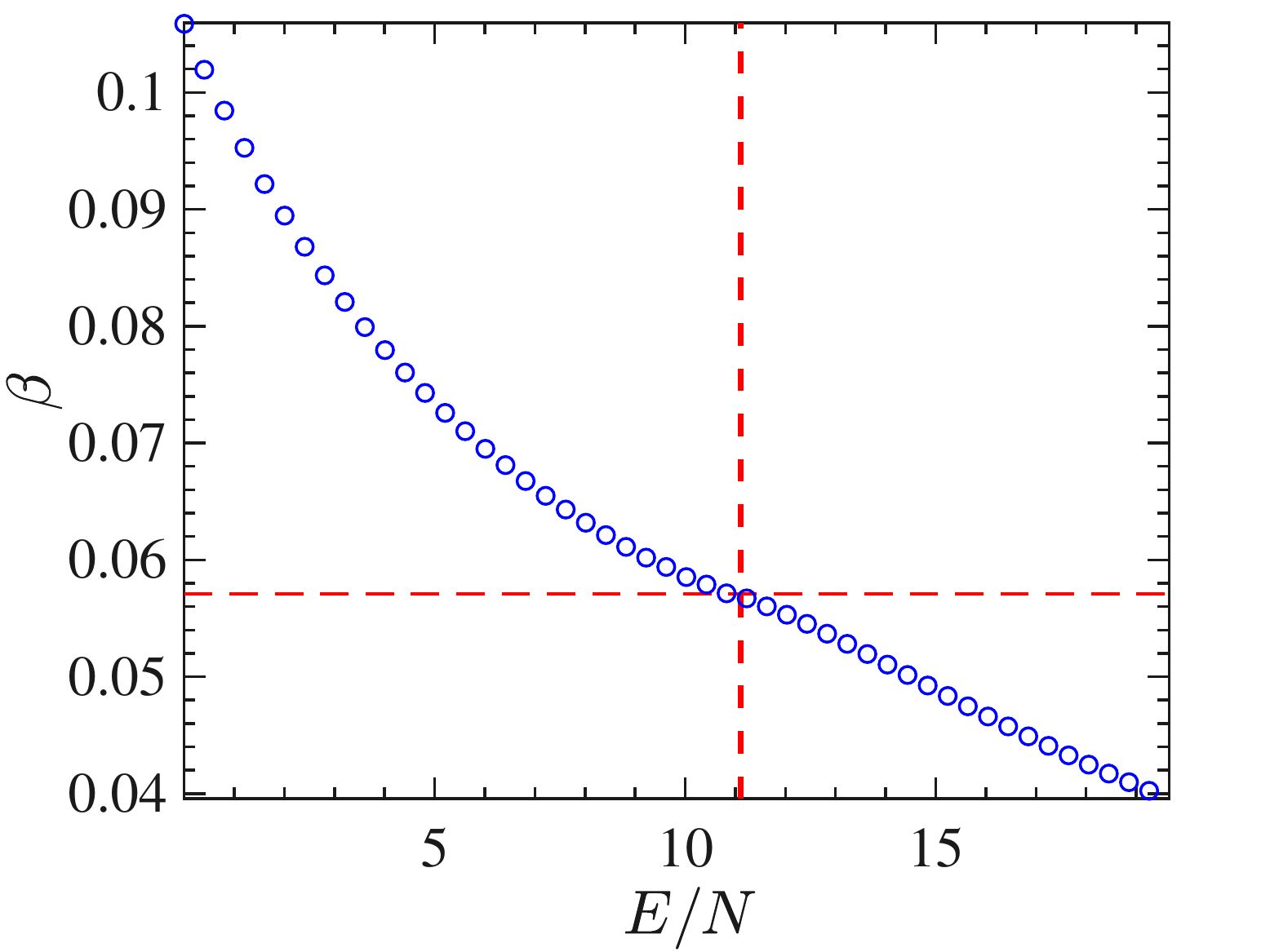}
\caption{\label{beta_e_2d}   $\beta$ vs $E/N$
obtained from the time average of $\Phi_1$ for several energies $E$ in
the case of the $2d$-$\phi^4$ model with $128 \times 128$ particles.}
\end{figure} 
The expected growth with the system-size, of the peak of the specific-heat in
correspondence of the phase-transition is shown in Fig. \ref{cv_e_2dn}. The curve
of the specific heat $C_V$ vs the energy-density $E/N$ has been obtained
via Eq. \eqref{eq:cv} where the averages have been again computed by means of time
averages of the quantities \eqref{Phi1} and \eqref{Phi2} according to Eq. \eqref{timeav}, for different lattice sizes, that is, 
 $24\times 24$ sites (open circles), $48\times 48$  sites
(open squares) and $128\times 128$  sites (crosses).
\begin{figure}[ht]
 \includegraphics[height=5.5cm]{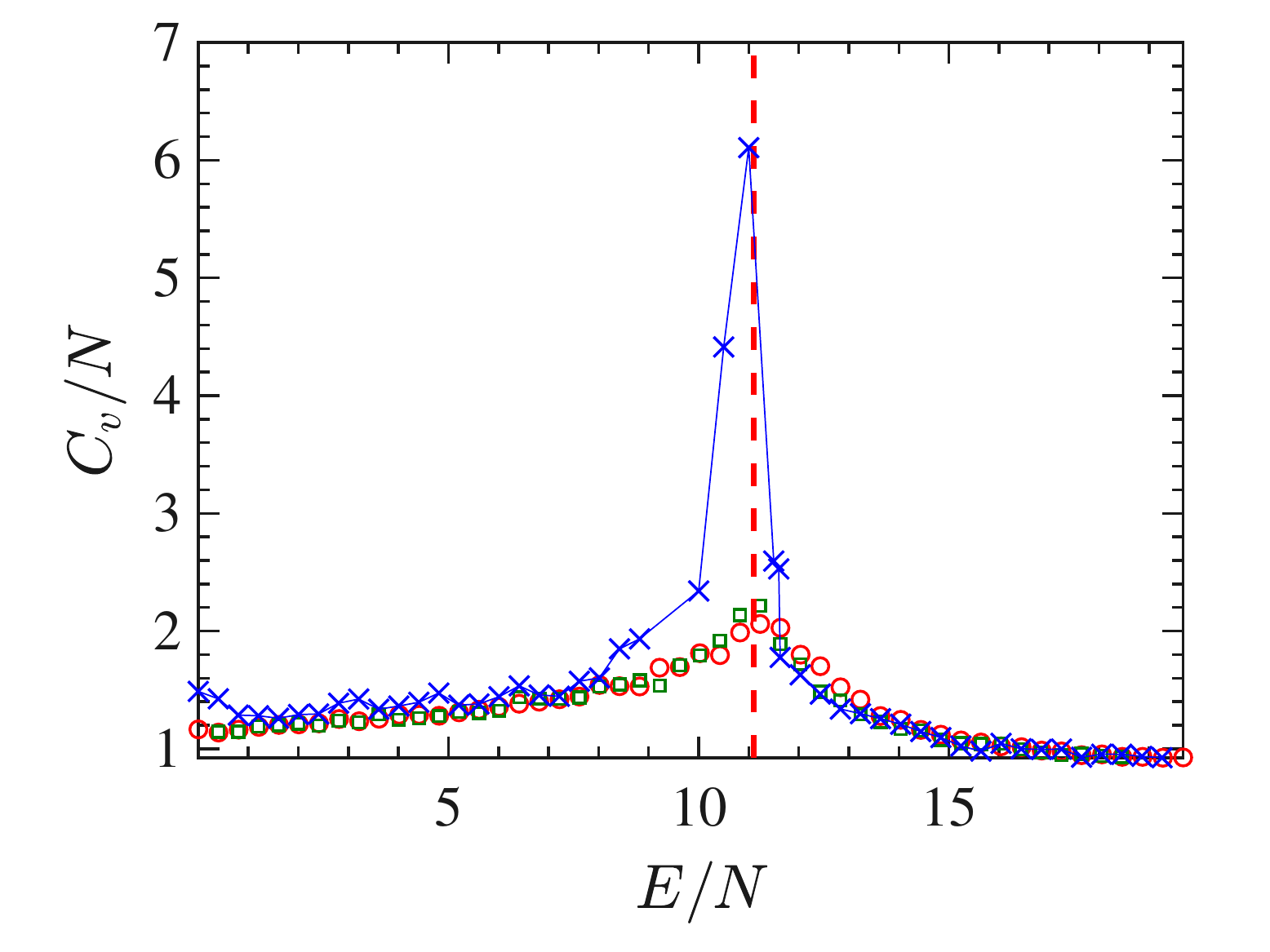}
\caption{\label{cv_e_2dn} The specific-heat per particle
$C_v/N$ is reported as a function of the energy density $E/N$ measured according to
Eq. \eqref{eq:cv} and performing time averages \eqref{timeav} of
the relevant quantities \eqref{Phi1} and \eqref{Phi2}.
The lattice sizes are: $24\times 24$ (open circles),
$48\times 48$ (open squares) and $128\times 128$ (crosses).}
\end{figure} 

Fig. \ref{d2s_e_2d} reports the second derivative of the
entropy with respect to the energy $E$. As mentioned above, the divergence of the specific heat stems from the vanishing of
this derivative.
This figure displays the outcomes of a numerical  derivation of the curve
$\beta(E)$ obtained for systems of different sizes: $24\times 24$ lattice
sites (open circles), and $48\times 48$ lattice sites (crosses). In addition, Fig.
\ref{d2s_e_2d} reports the values of $N \partial^2S/\partial E^2$ vs $E/N$
derived by means Eq. \eqref{d2sf} through time averages of $\Phi_1$ and
$\Phi_2$ in the case of a system with $24\times 24$ (open
squares), $48\times 48$ (full circles) and $128\times 128$  (stars)
lattice sizes. 
\begin{figure}[ht]
 \includegraphics[height=5.5cm]{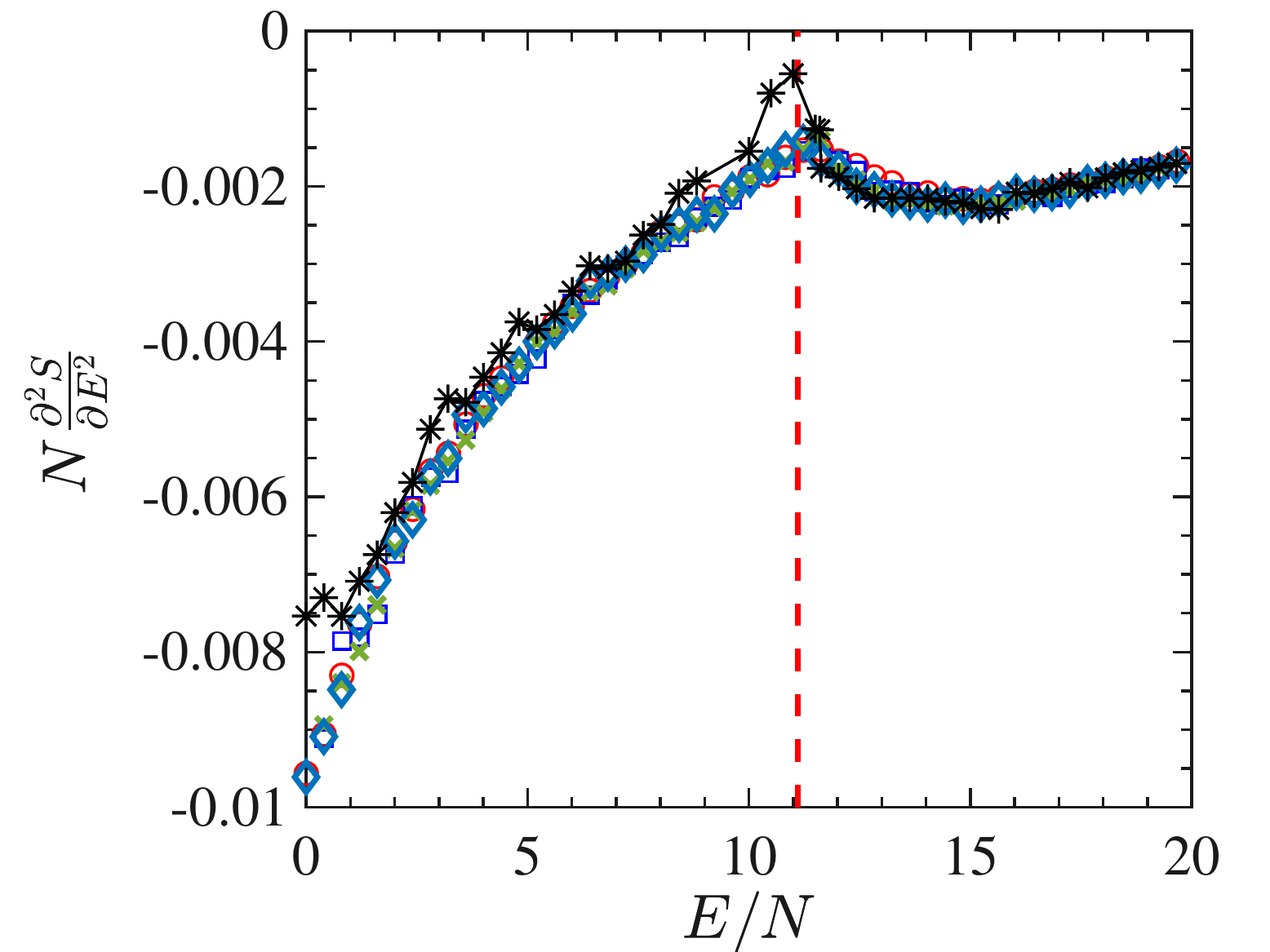}
\caption{\label{d2s_e_2d} $N \partial^2S/\partial E^2$ vs
$E/N$ derived with a numeric derivative of the curve $\beta(E/N)$.
The latter has been obtained as time average of $\Phi_1$ for
several values of the total energy $E$ in the case of a $24\times 24$
lattice (open circles) and a $48\times 48$ lattice (crosses).
Furthermore, the figure plots the $N \partial^2S/\partial E^2$
derived by the formula $N(\langle\Phi_2 \rangle- \langle \Phi_1\rangle^2 )$
in which the averages are temporal. Symbols refer to $24\times 24$ (open
squares), $48\times 48$ (full circles) and $128\times 128$  (stars)
lattice sizes, respectively. 
The figure shows distinctly the transition point, corresponding
to a discontinuity of the fourth order of the derivative of $S$.}
\end{figure} 
Remarkably, these numerical outcomes confirm that the growth with the system-size of
the specific-heat, in correspondence of the phase-transition, as a
consequence of the approaching of $N \partial^2S/\partial E^2$ to
zero as per Eq. \eqref{eq:cv0}. Fig. \ref{d2s_e_2d} shows that the larger  the system size the
closer  the value of $N \partial^2S/\partial E^2$ to zero, in correspondence of the
phase-transition point. 
\begin{figure}[h]
 \includegraphics[height=5.5cm]{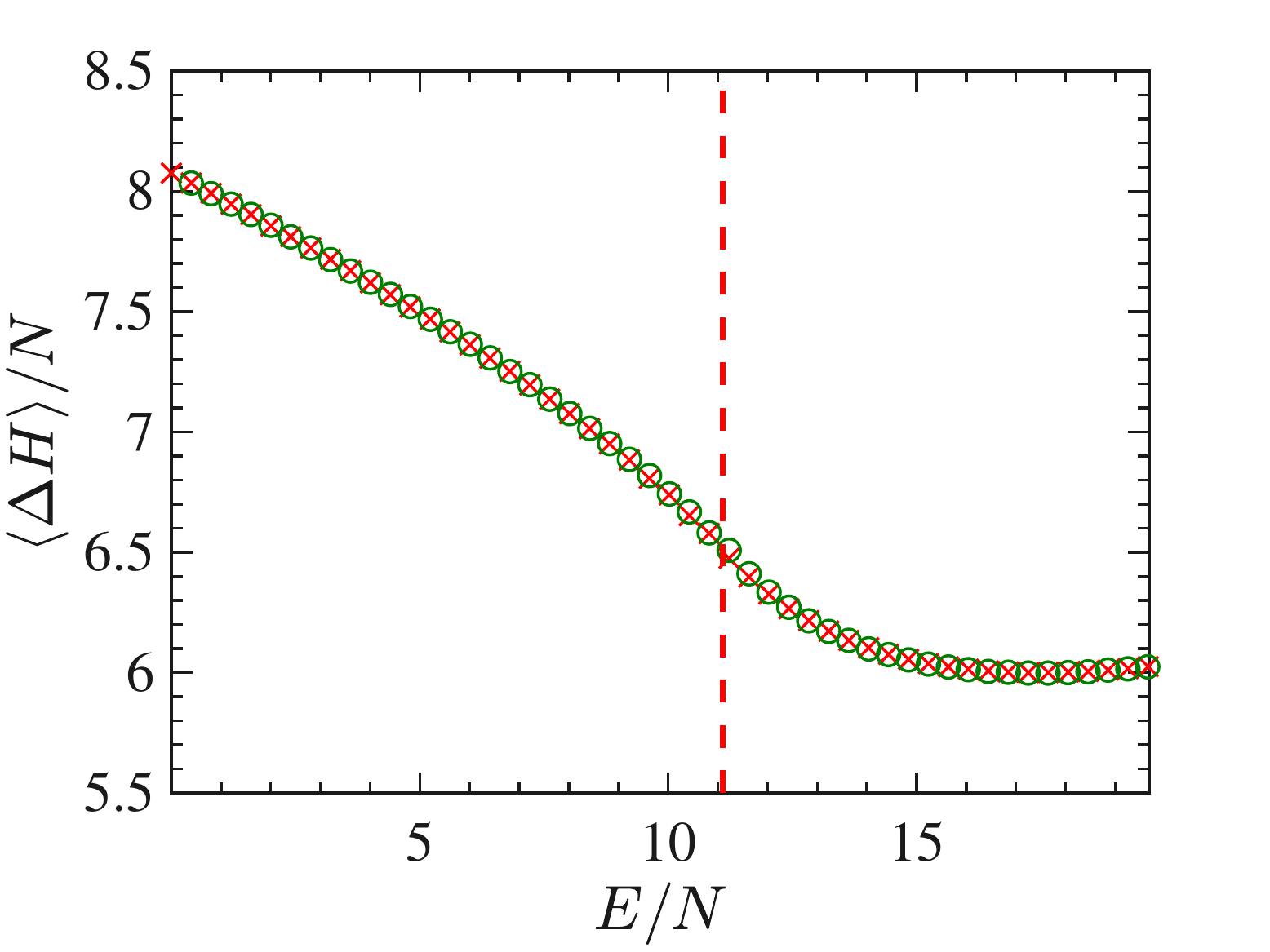}
\caption{\label{laplaH_e_2d} Figure report the time average of 
$\Laplace H/N$ as a function of $E/N$ in the case of a system with
$24\times 24$ lattice sites.}
\end{figure} 
In Figure \ref{laplaH_e_2d} the curve $\langle \triangle H\rangle/N$ vs $E/N$ is reported, that is  the time
average of the Laplacian of the Hamiltonian function per degree of freedom, and again it clearly shows an inflection point at the transition energy density. The quantity  $\langle \triangle H\rangle/N$ has a geometric meaning but of a different kind with respect to those related with the extrinsic curvature of the energy level sets. In fact,  as shown in the Appendix, it turns out that 
the Laplacian of the Hamiltonian [in Eq.\eqref{laplaKR}] coincides, apart from a constant, with the Ricci-curvature of a Riemannian manifold, an enlarged configurational space-time endowed with a metric due to Eisenhart \cite{Marcobook}. The geodesics of this manifold are just the natural motions of the Newton equations associated with the Hamiltonian of the system.

\subsection{Mean-field $\phi^4$ model}
In the present section, we analyse the results of the numerical simulations performed
for the mean-field $\phi^4$ model. This model
undergoes a second-order phase transition which
is displayed by the bifurcation of a standard order parameter, the magnetization, related with the breaking of the ${\mathbb Z}_2$ symmetry 
of the Hamiltonian of this system. This is shown in Figure \ref{order_e_mf} where the order parameter
$M=\langle {\cal M}  \rangle/N$, the average of the total
magnetization ${\cal M}$ in Eq.\eqref{M}, is reported as a function of the energy
density $E/N$. From this figure one can estimate the value $\epsilon_c \approx
25$ for the critical energy density corresponding to the phase transition.
\begin{figure}[h]
\includegraphics[height=5.5cm]{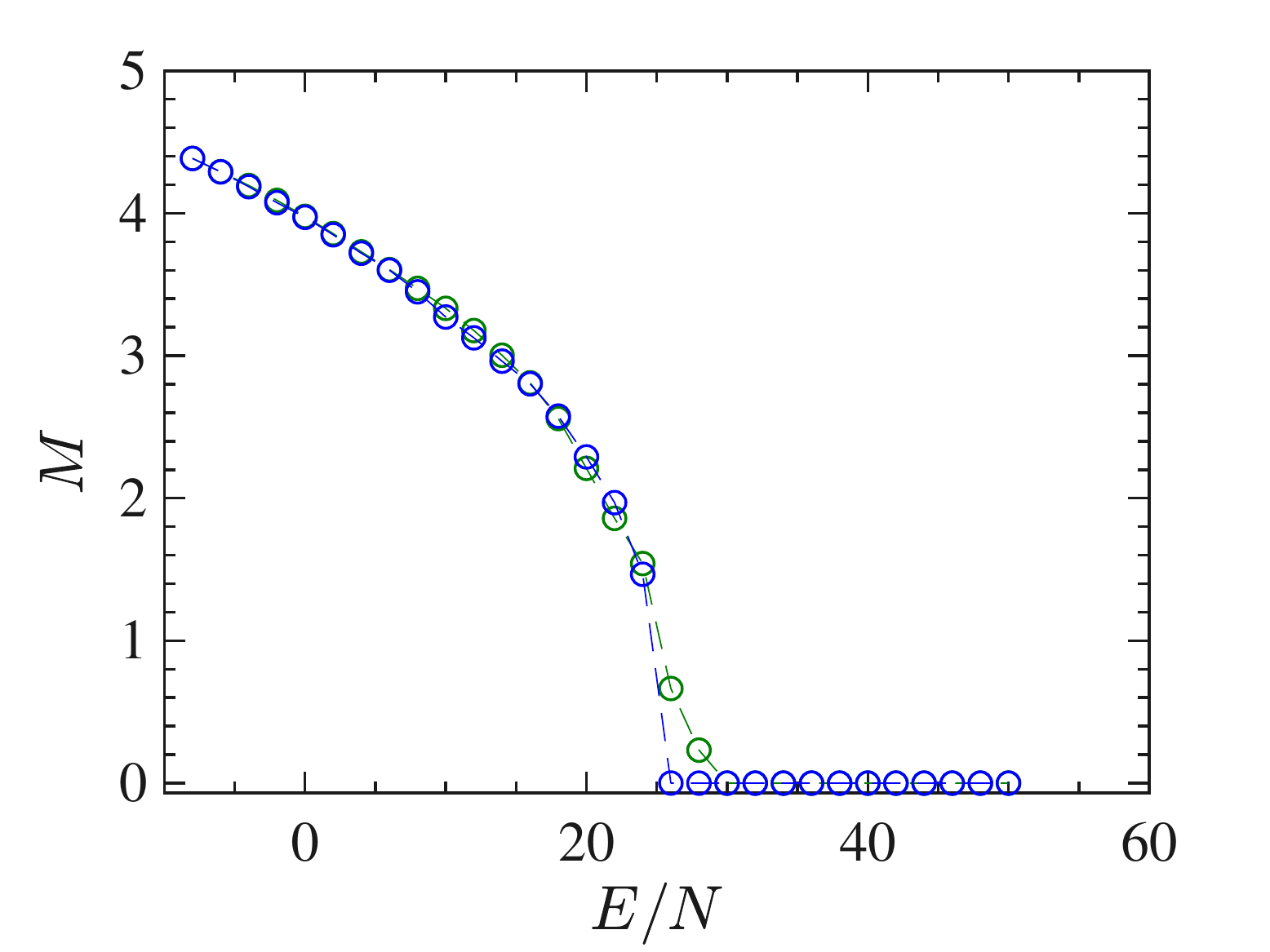}
\caption{\label{order_e_mf} The order parameter $M$ for the mean-field $\phi^4$ model is reported 
vs $E/N$ for $1024$ particles (green circles) and $2048$ particles (blue circles).}
\end{figure} 

With respect to the $2d$ model, the long-range interactions make this
system harder to simulate. In fact, considerable
difficulties have been encountered in computing stabilized time averages of the same quantities computed for the $\phi^4$ model with short-range interactions.
 These difficulties depend on 
the worsening of the properties of self-averaging of this model for energy values close to the
transition point, clearly due the long-range interactions.
Besides that, and again except for the order parameter, the mean-field model undergoes a phase transition which appears 
much "softer" than the one undergone by the $2d$ model.  This fact is put in evidence by the basic thermodynamic functions $T(E/N)$ and $\beta(E/N)$, computed though the time averages of $\Phi_1$ along with the numeric
integration of the equations of motion for different initial conditions,  and reported
in Figures \ref{e_T_mf} and \ref{beta_e_mf}, respectively. In particular the curve $\beta(E/N)$ does not display at all any feature to identify the presence of a 
transition. All in all, these functions are not very helpful neither to clearly identify the presence
of a phase-transition nor, possibly, its transition point. 
\begin{figure}[h]
 \includegraphics[height=5.5cm]{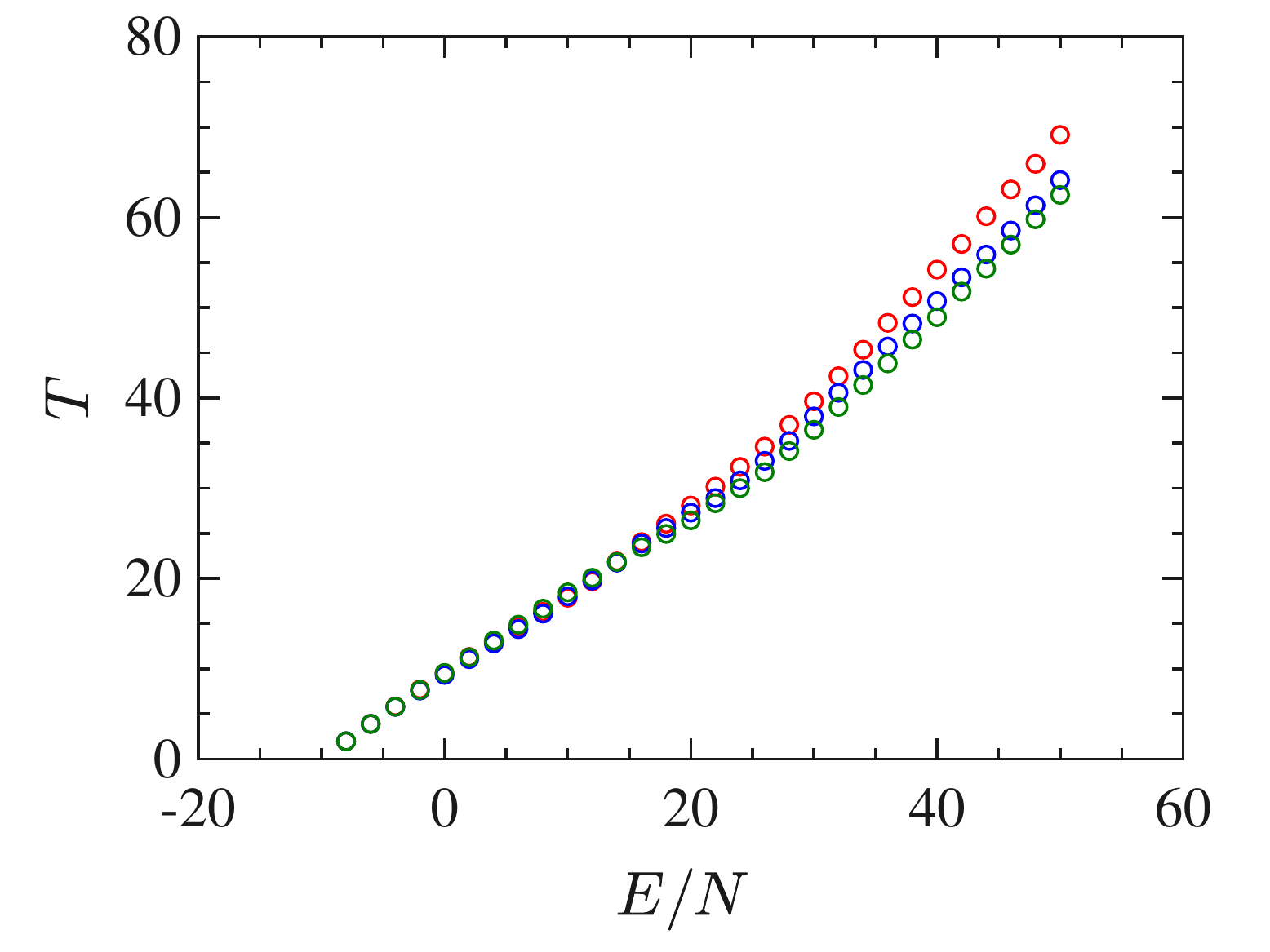}
\caption{\label{e_T_mf} $T$ vs $E/N$ for the mean-field $\phi^4$ model. 
$N=4096$ red circles, $N=2048$ blue circles, $N=1024$ green circles.}
\end{figure} 
\begin{figure}[h]
 \includegraphics[height=5.5cm]{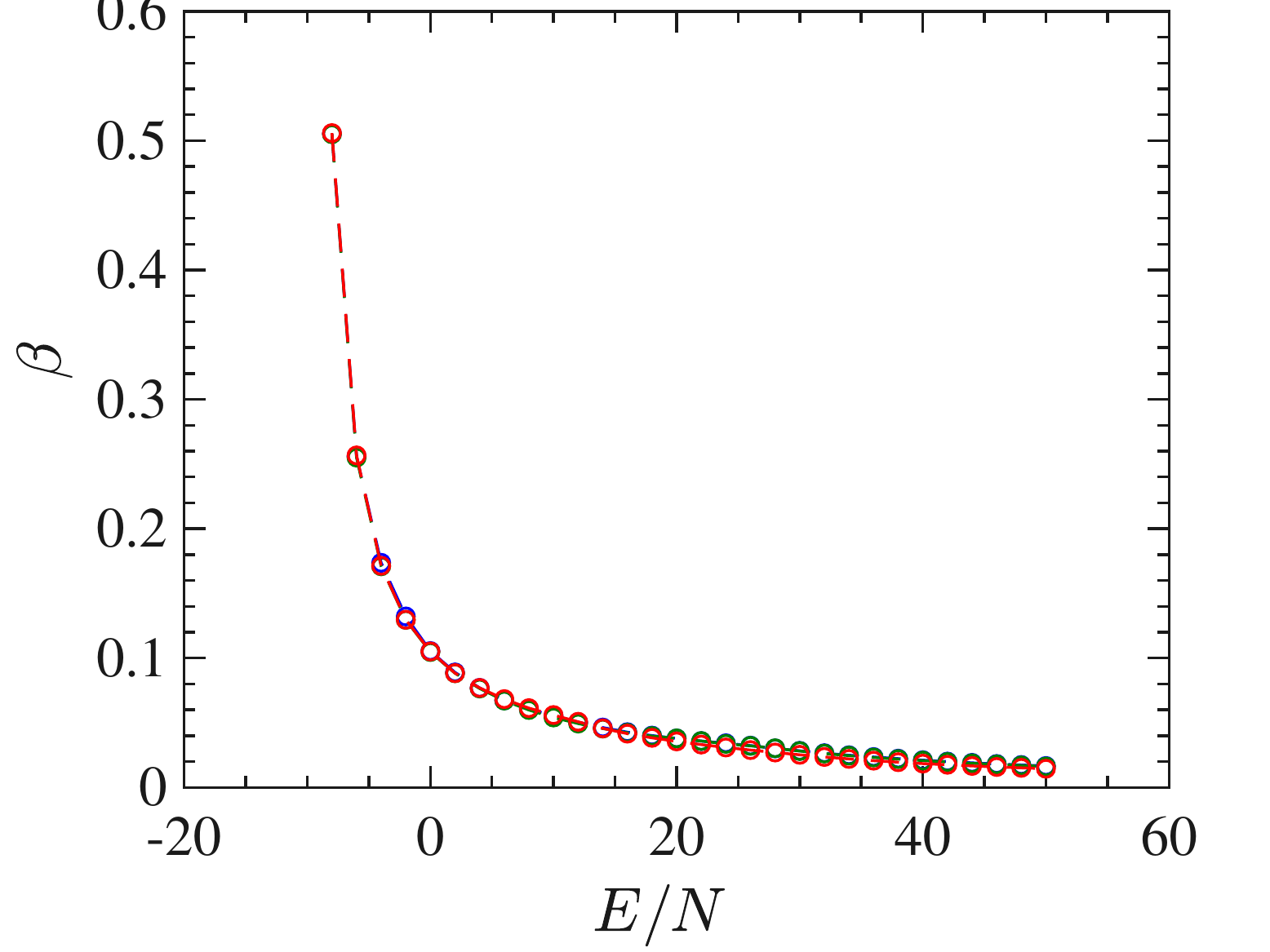}
\label{beta_e_mf}
\caption{The figure show the curve
$\beta$ vs $E/N$ for the mean-field $\phi^4$ model. 
$N=4096$ red circles, $N=2048$ blue circles, $N=1024$ green circles.}
\label{beta_e_mf}
\end{figure} 
In Figure \ref{d2s_e_mf} we report the derivative $N \partial^2 S/\partial E^2$ 
as a function of $E/N$ worked out in the same way as previously done for the short-range model.
The energy density pattern of this derivative is found to be very noisy, even after many millions of integration time steps, and this goes together with
a very bad outcome for the specific heat, which, on purpose, is not reported here.
To the contrary, and together with the order parameter, Figure \ref{laplaH_e_mf} shows an interesting  pattern of the time average of the Ricci curvature of the mechanical manifold $(M\times{\Bbb R}^2, g_e)$ (see Appendix) as a function of the energy density. The pattern of $\langle\Laplace H\rangle (E/N) /N$ displays a "cuspy" point in correspondence with the vertical red dashed line locating the phase transition point. Of course, within the obvious limits of numerical outcomes, such a "cuspy" point appears as an abrupt change of the second derivative of the Ricci curvature - with respect to the energy -  because above the transition point its pattern appears convex (of positive second derivative), whereas just below the transition point the values of the Ricci curvature appear to align along a straight segment, thus with a vanishing second derivative. Loosely speaking, this is reminiscent of similar jumps of the second derivative with respect to the energy of an average curvature function which has been found for a gauge model \cite{Pettini2019}.   

\begin{figure}[h]
\includegraphics[height=5.5cm]{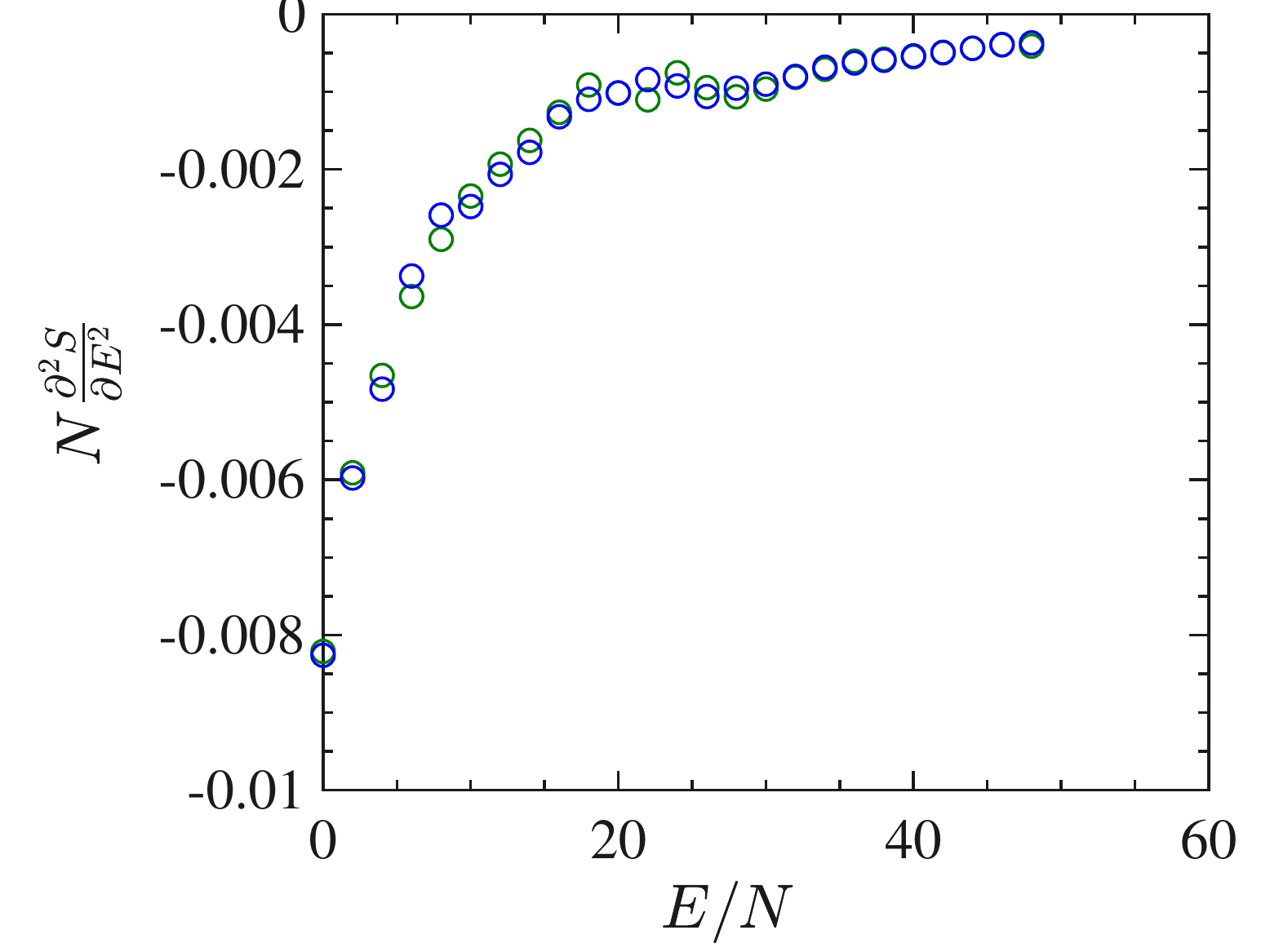}
\caption{\label{d2s_e_mf} The figure shows the plot
of the quantity $N \partial^2 S/\partial E^2$ 
vs $E/N$ derived with a numeric derivative of the curve
$\beta (E)$ for $1025$ particles.}
\end{figure} 

\begin{figure}[h]
 \includegraphics[height=5.5cm]{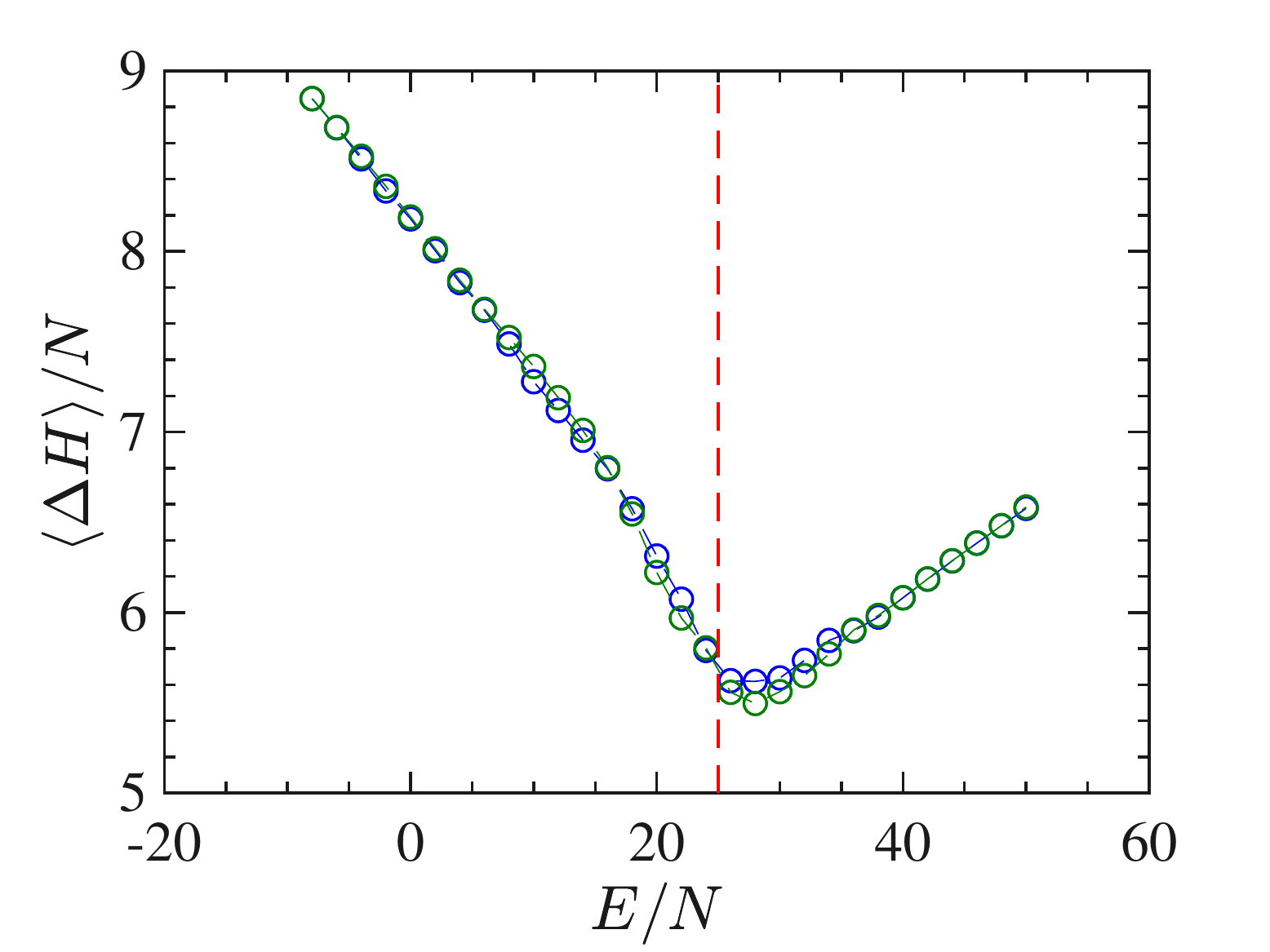}
\caption{\label{laplaH_e_mf} The figure shows the plot
of the quantity $\langle\Laplace H\rangle /N$
vs $E/N$ for $1024$ particles (green circles) and $2048$ particles (blue circles).}
\end{figure}

\section{Concluding remarks}
We have considered the second order phase transitions stemming from the same kind of ${\mathbb Z}_2$ symmetry-breaking phenomenon occurring in two $\phi^4$ models.
Besides the standard detection of the presence of a phase transition through the bifurcation of an order parameter, we have focused on basic geometric properties of different manifolds, highlighting that the values of thermodynamic observables, like temperature and specific heat, and their functional dependence on the energy are the consequences of more fundamental changes with energy of curvature properties of the energy level sets in phase space. The conceptual interest of this fact is that a phase transition phenomenon can be seen as just depending on the interaction potential of the forces acting among the degrees of freedom of a system, that is, the possibility for a system of undergoing a phase transition is already "encoded" in its Hamiltonian function and thus can be read in the variation of some extrinsic curvature properties  of the hypersurfaces $H(p,q)=E$ foliating the phase space. When the variations with energy the geometry of these level-set manifolds are too "mild", as is the case of the mean-field $\phi^4$ model, one can again recover a rather sharp geometric signature of the transition by considering the energy variation of the Ricci curvature of a manifold the geodesics of which are the motions of the system. In other words, in both cases, a phase transition phenomenon can be seen as stemming from a deeper level than the usual one which consist of attributing them to a loss of analyticity of the statistical measures in the thermodynamic limit.
The statistical measures represent an "epistemic" description of the occurrence of phase transitions, in what statistical measures do not correspond to physically measurable entities, whereas the forces acting among the degrees of freedom of a system belong to an "ontic" level because forces are real physical entities, velocities of the kinetic energy and potentials can be in principle measured, so that for an energy conserving closed system the quantities entering the relation $H(p,q)=E$ are real physical ones.

Finally,  since geometric indicators, like the Ricci curvature, are independent of the order parameter among the other thermodynamic quantities, the proposed geometric analysis can be applied also in the case of systems that undergo phase-transitions in absence of a global symmetry breaking and thus in the absence of an order parameter \cite{Pettini2019}.

\begin{acknowledgments}

R. Franzosi thanks the support by the QuantERA project 
“Q-Clocks” and the European Commission.
\end{acknowledgments}

\appendix

\section{Eisenhart Metric on Enlarged Configuration Space-Time $M\times{\Bbb R}^2$}
\label{Eisenhart_metric}
The natural motions of a standard Hamiltonian system, that is, having a quadratic kinetic energy term, can be identified with a geodesic flow on a Riemannian manifold.
Among the other possibilities, Eisenhart proposed a geometric formulation of Hamiltonian/Newtonian dynamics by resorting to
an enlarged configuration space-time
$M\times {\Bbb R}^2$ having the local coordinates
$(q^0,q^1,\ldots,q^i,\ldots,q^N,q^{N+1})$. This space can be
endowed
with a nondegenerate pseudo-Riemannian metric \cite{Eisenhart} whose
arc length is
\begin{equation}
\begin{split}
ds^2 &= \left( g_e \right)_{\mu\nu}\, dq^{\mu}dq^{\nu} = \\
&~~~~~~a_{ij} \, dq^i dq^j -2V(q)(dq^0)^2
+ 2\, dq^0 dq^{N+1} ~,
\label{g_E}
\end{split}
\end{equation}
where $\mu$ and $\nu$ run from $0$ to $N+1$ and  $i$ and $j$ run from 1 to $N$.
The following theorem \cite{Marcobook} holds
\smallskip

\noindent{\bf Theorem (Eisenhart)}
\textit{
The natural motions of a Hamiltonian dynamical system
are obtained as the canonical projection
of the geodesics
of $(M\times {\Bbb R}^2,g_e)$ on the configuration
space-time,
$\pi : M\times {\Bbb R}^2 \mapsto M\times {\Bbb R}$.
Among the totality of geodesics, only those whose
arc lengths are
positive definite and are given by
\begin{equation}
ds^2 = c_1^2 dt^2
\label{ds_Eisenhart}
\end{equation}
correspond to natural motions; the
condition (\ref{ds_Eisenhart}) can be equivalently cast in the following
integral form
as a condition on the extra coordinate $q^{N+1}$:
\begin{equation}
q^{N+1} = \frac{c_1^2}{2} t + c^2_2 - \int_0^t { L}\,
d\tau~,
\label{qN+1}
\end{equation}
where $c_1$ and $c_2$ are given real constants.
Conversely, given a point $P \in M \times {\Bbb R}$ belonging
to a trajectory
of the system, and given two constants $c_1$ and $c_2$, the
point
$P' = \pi^{-1} (P) \in M \times {\Bbb R}^2$, with
$q^{N+1}$ given by
(\ref{qN+1}), describes a geodesic curve in $(M\times {\Bbb R}^2,g_e)$
such that $ds^2 = c_1^2 dt^2$.}

\smallskip

\noindent The explicit table of the entries of the Eisenhart metric is 
\begin{equation}
g_e = \left(
\begin{array}{ccccc}
-2V(q)& 0       & \cdots        & 0     & 1     \\
0       & a_{11}& \cdots        & a_{1N}& 0     \\
\vdots  & \vdots& \ddots        & \vdots& \vdots\\
0       & a_{N1}& \cdots        & a_{NN}& 0     \\
1       & 0     & \cdots        & 0     & 0     \\
\end{array} \right)\ , \label{gE}
\end{equation}
where $a_{ij}$ is the kinetic energy metric. The only 
non vanishing Christoffel symbols, for $a_{ij} =
\delta_{ij}$, are 
\begin{equation}
\Gamma^i_{00} = - \Gamma^{N+1}_{0i} = \partial_i V~,
\label{Gamma_E}
\end{equation}
whence the geodesic equations 
\[
\frac{d^2 q^i}{ds^2}+ \Gamma^i_{jk}\frac{dq^j}{ds}\frac{dq^k}{ds}=0\ ,
\]
reduce to
\begin{eqnarray}
\frac{d^2q^0}{ds^2}  &=&  0 ~, \label{eqgeo0}\\
\frac{d^2q^i}{ds^2} +\Gamma^i_{00}
\frac{dq^0}{ds}\frac{dq^0}{ds}  &=&  0\ ,
\label{eqgeoi}\\
\frac{d^2q^{N+1}}{ds^2} +\Gamma^{N+1}_{0i} \frac{dq^0}{ds}
\frac{dq^i}{ds}  &=&  0 \ ; \label{eqgeoN+1}
\end{eqnarray}
using $ds = dt$ one obtains
\begin{eqnarray}
\frac{d^2q^0}{dt^2}  &=&  0\ , \label{eqgeo0t}\\
\frac{d^2q^i}{dt^2}  &=&  - \frac{\partial V}{\partial q_i}
~,\label{eqgeoit}\\
\frac{d^2q^{N+1}}{dt^2} & =&   - \frac{d{ L}}{dt} ~.
\label{eqgeoN+1t}
\end{eqnarray}
Equation (\ref{eqgeo0t}) states only that $q^0=t$.
The $N$ equations (\ref{eqgeoit})
are Newton's equations, and (\ref{eqgeoN+1t}) is
the differential version of (\ref{qN+1}).

The Riemann curvature tensor, associated with Eisenhart metric, has the following nonvanishing components
\begin{equation}
R_{0i0j} = \partial_i\partial_j V~;
\end{equation}
thus the only nonzero component of the Ricci tensor is
\begin{equation}
R_{00} = \triangle V~,
\label{ricci_eisenhart}
\end{equation}
finally the Ricci curvature is
\begin{equation}
K_R(q,\dot q)=R_{00}\dot q^0\dot q^0\equiv \triangle V\ ,
\label{eqKR}
\end{equation}
so that $\triangle H/N$ is just
\begin{equation}
\frac{\triangle H}{N} = \frac{K_R(q,\dot q)}{N} + 1 \ .
\label{laplaKR}
\end{equation}

\bibliography{references}

\end{document}